# Experimental Evidence of Direct Exchange Interaction Mediating Intramolecular Singlet Fission in Weakly-Coupled Dimers

Oskar Kefer,[1] Pavel V. Kolesnichenko,[1] Lukas Ahrens,[2] Jan Freudenberg,[2] Uwe H. F. Bunz[2] and Tiago Buckup[1*]

[1] Physikalisch-Chemisches Institut, Ruprecht-Karls-Universität Heidelberg, D-69120 Heidelberg, Germany.

[2] Organisch-Chemisches Institut, Ruprecht-Karls-Universität Heidelberg, D-69120 Heidelberg, Germany.

* Corresponding Author:

    tiago.buckup@pci.uni-heidelberg.de

**Abstract**

The electronic interaction between an optically active singlet state ($S_1S_0$) and a dark state of singlet multiplicity, known as correlated triplet pair ($^1$[TT]), plays a crucial role in the effective transformation from $S_1S_0$ to $^1$[TT] during intramolecular singlet fission (iSF). This process is understood through mechanisms such as direct exchange coupling and incoherent processes that involve super-exchange coupling through charge-transfer states. However, most insights into these mechanisms are derived from theoretical studies due to the difficulties in obtaining experimental evidence. In this study, we investigate the excited-state interactions between $S_1S_0$ and $^1$[TT] in spiro-conjugated iSF sensitizers by employing transient two-dimensional electronic spectroscopy. This approach allows us to focus on the early stages of the conversion from $S_1S_0$ to $^1$[TT]. Upon optical excitation, a superposition of $S_1S_0$ and $^1$[TT] is created, which gradually transitions to favor $^1$[TT] within the characteristic time frames of iSF. The observed high-order signals indicate circular repopulation dynamic that effectively reinitiates the iSF process from higher energy electronic states. Our findings, supported by semi-quantum-mechanical simulations of the experimental data, suggest the presence of a direct iSF mechanism in the dimers, facilitated by weak non-adiabatic coupling between $S_1S_0$ and $^1$[TT]. This experiment provides new insights into the equilibrium between the two electronic states, a phenomenon previously understood primarily through theoretical models.

## 1. Introduction

Conversion of one excitation into two charge carriers via singlet fission (SF) offers a promising strategy for reducing thermal losses during solar energy conversion.[1-4] Its unique quality of producing high-spin electronic states has also raised interest for dynamic nuclear polarization.[5-9] SF is realized by the transformation of one optically bright singlet excited state ($S_1$) into two long-lived triplet states ($T_1$), which is facilitated by a number of intermediaries, such as the correlated triplet pair states ($^x[TT]$, $x = 1, 3, 5$), as summarized by the general mechanism[10]

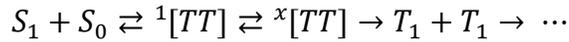

$$S_1 + S_0 \rightleftarrows {}^1[TT] \rightleftarrows {}^x[TT] \rightarrow T_1 + T_1 \rightarrow \cdots$$

The reaction can unfold on ultrafast timescales[11-14] due to spin-conserving internal conversion of $S_1$ to a triplet-pair. In this process, the spins of the individual triplets are initially correlated, resulting in an overall singlet multiplicity ($^1[TT]$).[15-16] Eventually, higher spin-states ($x = 3,5$) start playing a role in the process,[7-9, 15] and the correlation dephases, forming individual triplets ($T_1$). Following this description, it is clear that the initial transition forming the $^1[TT]$ is crucial for the outcome of the whole process, capturing the attention of several works in recent decades. Previous experimental[17-23] and theoretical[8, 24-30] studies have unveiled comprehensive details regarding the different pathways that contribute to the generation of this elusive multiexcitonic state. One such pathway involves a single-step coherent transition of $S_1+S_0$ ($S_1S_0$ in covalently bound dimers) to the $^1[TT]$ state, in which the de-excitation of the excited singlet promptly promotes a ground-state chromophore to the triplet manifold. In this case, the rate and efficiency of the conversion is governed by the coupling of the adiabatic electronic states $S_1$ and $^1[TT]$. Other contributions to SF involve incoherent pathways that follow a sequential exchange of excitation energy. These processes typically include intermediate charge-transfer (CT) states, in which the charges can separate between the two chromophores, or a superposition of different electronic states mediated by super-exchange coupling.[22, 31] However, if chromophores and environment favor charge separation, the CT state may decrease in energy to a point where it traps the excitation and inhibits SF to proceed further.[32-34] Thus, a paramount goal of SF research has been the identification and quantification of the different interaction pathways to be able to make predictions on the efficiency of the fission process (and optimize alongside it) prior to introduction of chemical modifications.

Typically, the electronic states involved in singlet fission (SF) are analyzed using transient absorption (TA) spectroscopy. This technique measures the perturbation in the system's absorption caused by an intense pump pulse. In many instances, it enables the differentiation between the intermediate steps of SF. [11, 18, 21, 32, 34-46] Analysis of time-resolved dynamics quantifies relevant SF-characteristics, such as time scales and quantum yields of triplets. Further insight into the microscopic mechanism of $^1$[TT]-formation is gained by, e.g., subjecting the chromophores to different chemical environments (e.g., via solvent polarity)[18, 43, 47-49] due to their profound influence in (de-)stabilizing the charge-separated CT states. Conversely, SF characteristics that are unaffected by the chemical environment point toward direct SF.

While the information content of TA is plentiful to study and understand SF, it lacks insight into the association of states. Two-dimensional electronic spectroscopy (2DES), on the other hand, can reveal correlations between electronic states and has been instrumental in the study of a multitude of materials, ranging from biological light-harvesting complexes in photosynthesis to semiconductors designated for photovoltaic devices.[18, 50-82] This is achieved by the additional measurement of the ultrafast oscillations caused by electronic coherences that are introduced during the interaction with the laser electric fields. Fourier transformation with respect to the corresponding time axis of such coherences disperses the spectral information captured by the probe spectrum along an additional dimension, the excitation axis $\lambda_1$, thus establishing correlations between excitation and emission events. Few 2DES studies of SF highlight the interplay of vibrational modes in the interaction of electronic states. In particular, *Bakulin* et *al.*[75] proposed the ultrafast SF (<100 fs) in bulk pentacene is mediated by strong vibronic coupling that leads to electronic-state mixing between $S_1$ and $^1[TT]$. This, in turn, increases the oscillator strength for an otherwise dark ground-state transition $S_0 \rightarrow {}^1[TT]$, enabling even a direct observation of the multiexcitonic state. Similarly, *Mandal* et *al.*[22] reported on the SF in terrylenediimide dimers, for which the dominant process mediating SF in non-polar solvents was proposed to be the result of electronic state mixing mediated by vibronic coupling. In contrast to pentacene, considerable contributions from CT-states with nearly degenerate energy levels to $S_1$ and $^1[TT]$ were observed in the weakly coupled system. According to *Wang* et *al.*,[66] vibronic mixing also plays a decisive role in endothermic SF of crystalline tetracene. In this case, mixing of high-energy $S_1$-vibrational modes with energetically higher $^1[TT]$ states leads to the excitation-energy-independent SF dynamics.

The experimental challenges of studying all aspects of SF with 2DES or other third-order non-linear spectroscopies (such as TA) become apparent from the prior literature examples. This is due to the optically dark transitions of $S_0 \rightarrow {}^1[TT]$ (or CT-states), which do not allow for a direct interrogation of correlations among the relevant excited states, unless SF is mediated by strong (vibronic) coupling (*vide supra*). Such challenges were recognized by various groups[83-88] reporting on excited-state processes that include, among others, optically dark CT-states. This apparent limitation is possible to overcome with transient 2DES, which can measure higher-order nonlinearities.[85, 89] In transient 2DES, the high-order responses are typically measured by preparation of the excited-state with an actinic pump prior to interrogation with the usual 2DES pulse-sequence. For example, *Mandal* et *al.*[83] have shown that by using transient 2DES it is possible to interrogate the CT process in a donor-acceptor system of perylene and perylenediimide at arbitrary points along the reaction coordinate with the possibility to reintroduce vibrational coherences at later stages of the reaction.

In this study, we use transient two-dimensional electronic spectroscopy (2DES) to directly investigate, for the first time, the excited-state correlations during singlet fission (SF) in a series of spiro-conjugated homodimers.[90] The general SF process for this type of dimers has been presented in our earlier works,[35-36] which showed the linker-defined SF characteristics. In this study, once again we focus our attention on the linker moiety in regard to SF in TIPS-Tetra-Aza-Pentacene (TAP) dimers: a series of different linkers bridging the two identical TAP chromophores are considered. However, more attention is put towards the initial step of SF, which comprises the conversion of $S_1$ to $^1[TT]$. We are able to fully capture high-order SF dynamics, unveiling an intricate interplay of high-energy electronic states. Simulations of transient 2DES correlation maps in accordance with our experimental results quantify the observed excited-state correlations between the singlet and correlated triplet-pair states, for which, SF is mediated by direct exchange coupling.

The following sections will give a brief description of the experimental method, explaining the most fundamental aspects and necessary information to intuitively understand experimental results. Afterwards, experimental results of 2DES and transient 2DES are presented and discussed in light of theoretical simulations, quantifying our observations.

## 2. Transient Two-Dimensional Electronic Spectroscopy

The experimental approach for transient 2DES was described in detail elsewhere[89] and is briefly outlined in the Supplementary material together with the theoretical framework for data simulation. In general, the experiment and simulation follow the pulse sequence shown in Figure 1a. Interaction of the sample with the incident electric fields leads to coherence ($\tau$ and t) or population ($T_{AP}$ and T) dynamics, depending on the state of the system and the number of interactions with the incident electric fields.[91-93]

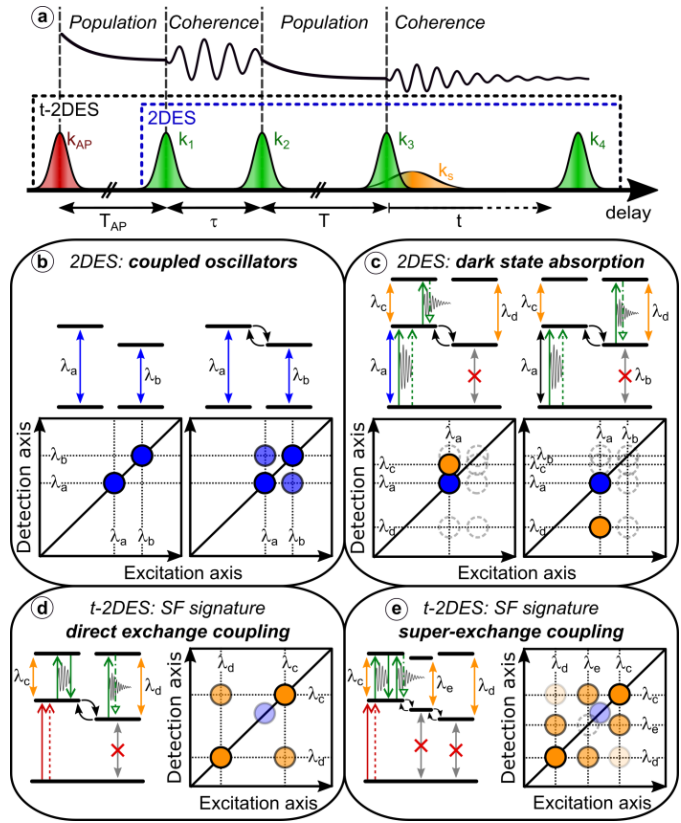

*Figure 1:* Introduction to (transient) 2DES. a) Pulse sequence of (transient) 2DES (bottom, see text for more details). System evolution between the pulses is shown at the top. b)-e) Simple exemplary models (top) used to demonstrate expected peak positions (blue and orange circles) in the corresponding 2DES correlation maps (bottom). The peak positions are related to the indicated transition wavelengths $\lambda_x$ (x=a,b,c,d). Coherence time periods are indicated in c)-e) as oscillating traces emerging from the respective electric field interaction. Exchange coupling between states is indicated by bent arrows. Forbidden transitions are indicated by red crosses. The traced peaks in c),e) correspond to expected peak positions of the dark-state transitions.

In the case of transient 2DES, a portion of the ground-state population (later denoted as $S_0S_0$) is shifted towards the excited state by an actinic pump (AP, with the wavevector $k_{AP}$), and after a delay, coined as actinic pump delay ($T_{AP}$), the interaction with the first pulse ($E_1$, with $k_1$) of the 2DES pulse sequence takes place. The rapidly oscillating coherence induced by $E_1$ is subsequently populated by a second pulse ($E_2$, with $k_2$). The delay between $E_1$ and $E_2$ defines the coherence time $\tau$, and the evolution afterwards develops in the population time (T) domain. Interaction with the third pulse ($E_3$, with $k_3$) leads to the final coherence that stimulates the signal ($E_s$, with $k_s$), which is detected interferometrically with a local oscillator ($E_4$, with $k_4$).[94-98]

The individual measurement of electronic coherences set by $E_1$ and $E_3$ (and evolving during $\tau$ and t, respectively) is the key methodology of 2DES: the spectral information of the stimulated signal in the detection frequency domain ($\lambda_3$) is extended into the excitation frequency domain ($\lambda_1$), thus correlating such coherences in the 2D domain (spanned by $\lambda_1$ and $\lambda_3$). For illustration purposes, sketches of 2D correlation maps for different model systems are shown in Figure 1b-e. For two non-interacting excited states with a common ground state (Figure 1b), two peaks at $\{\lambda_a|\lambda_a\}$ ({excitation|detection} wavelengths) and $\{\lambda_b|\lambda_b\}$ can be differentiated along the diagonal in the 2D map. In this case, equal coherence oscillations during $\tau$ and t are measured. Introducing an exchange coupling between the excited states results in two additional peaks located in the cross-peak regions $\{\lambda_a|\lambda_b\}$ and $\{\lambda_b|\lambda_a\}$. They correlate the excitation of state **a** (with $\lambda_a$) with the detection of state **b** (with $\lambda_b$) and vice versa, revealing the interaction. This is one of the main advantages of 2DES and extension along the excitation axis: interstate coupling can be distinguished by cross-peaks, which would otherwise be obscured in one-dimensional experiments (such as TA spectroscopy) where the detected signal typically appears integrated along the excitation wavelength axis.

When an optically dark state is included in the exemplary model (Figure 1c), as is commonly the case for singlet fission with $^1$[TT] (and CT-states), 2DES may also be insensitive towards certain correlations resulting in missing peaks in the 2D correlation map (Figure 1c). The existence of dark states may become evident from the emergence of their delayed (in the population time domain) excited-state absorption (ESA) features {$\lambda_a|\lambda_d$}. They appear at the same excitation wavelength $\lambda_a$ as that of the initially populated excited state (with ESA at {$\lambda_a|\lambda_c$}) thus correlating various excited-state absorption features (arising from the excitation of the ground state) to an optically bright electronic state, e.g., $S_1S_0$ in the case of SF. Especially for SF, this complicates experimental evaluation of the intricate interaction between $S_1S_0$ and $^1$[TT].

Pre-excitation of the system with an AP followed by the 2DES pulse sequence establishes correlations between excited states, as exemplified in the energy diagrams of Figure 1d,e. Certain interaction diagrams correlate the measured coherence oscillations (during $\tau$ and t) that can be attributed solely to coherences of excited electronic states. For direct and CT-mediated SF pathways introduced above, spectral peaks like those shown in the 2D maps of Figure 1d,e can therefore be anticipated. The direct coupling elements of $S_1S_0$ and $^1$[TT] (Figure 1d) could be distinguished by cross-correlation signals located around {$\lambda_c|\lambda_d$} and {$\lambda_d|\lambda_c$}, following the reasoning similar to the example shown in Figure 1b. The involvement of virtual or real CT-states that mediate SF via super-exchange coupling (Figure 1e) should be recognizable by additional cross-peaks beyond those observed for the direct coupling with missing/reduced cross-correlations of $S_1S_0$ and $^1$[TT].

## 3. Results

**Singlet Fission Signatures of TAP Spiro-conjugated Dimers**

Presentation of results will focus on one of the spiro-conjugated homodimers in tetrahydrofuran (THF) as solvent, unless noted otherwise, due to the overarching similarity in the spectra of the dimers that bear the same chromophore (TAP) and the lack of solvent-dependent intramolecular singlet fission (iSF) characteristics. The results that are not shown here can be found in the Supplementary material. The relevant SF-related spectral signatures of the TAP spiro-conjugated dimer are summarized in Figure 2. The ground-state absorption spectrum (Figure 2a) features three distinct vibronic absorption bands at 683, 625 and 580 nm. The AP spectrum is tuned to overlap the ground-state absorption centered around $\lambda_{S0(1)} = 625$ nm (corresponding to the transition $S_0S_0(\nu_0) \rightarrow S_1S_0(\nu_1)$). The 2DES sequence spectrum is tuned to be resonant with the excited-state transitions of the $S_1S_0$ ($\lambda_{S1} = 593$ nm) and $^1$[TT] ($\lambda_{1TT} = 550$ nm), as shown in Figure 2b (see Supplementary material for further details) and overlaps the ground-state absorption band at 580 nm ($\lambda_{S0(2)} = 580$ nm). Molecular transitions that are driven by the electric fields of the AP and 2DES sequence are summarized in the diagram in Figure 2c. The relative energies of the eigenstates in Figure 2c are determined from the optical spectra given in Figure 2a-b and theoretical results presented in Ref. [35].

Spectral correlation maps recorded with the conventional 2DES (without AP) are presented in Figure 2d and capture the conversion of the $S_1S_0$ to $^1$[TT], based on the assignment of the observed peaks to the related transitions in Figure 2c. In particular, the pronounced negative features located around {580|592 nm}, that are present for early population times (T < 80 ps), are assigned to the ESA of $S_1S_0$, relating it to the excitation of the ground-state into a higher vibrational level of $S_1S_0(v_2)$. The $S_1S_0$ features decay with the time constant of 74±4 ps giving rise to an ESA centered around {580|550 nm}, which is assigned to the $^1[TT]$. The timescale of the initial SF process captured with 2DES is in excellent agreement with the SF time constant acquired with TA after excitation of the 580-nm absorption band (see Supplementary material).

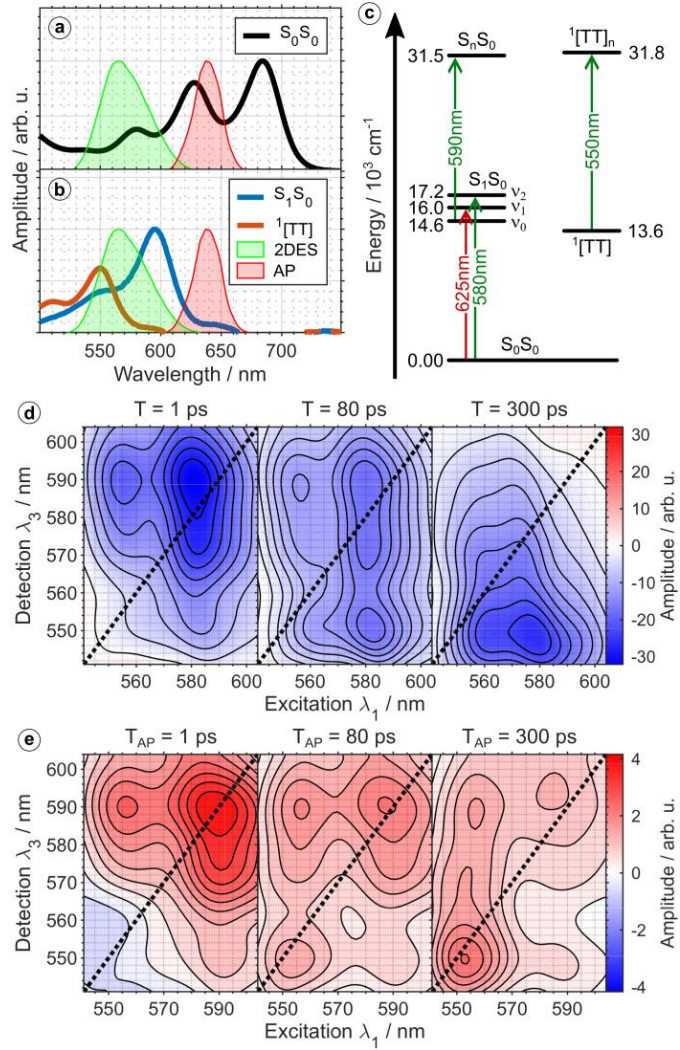

*Figure 2: Singlet fission in TAP spiro-conjugated dimers. a) and b) show the ground- ($S_0S_0$) and excited-state ($S_1S_0$ and $^1[TT]$) absorption spectra (see Supplementary material for more information), respectively, compared to the actinic pump and 2DES sequence spectra. c) Energy diagram of relevant eigenstates and vibrational states $v_i$ of $S_1S_0$. Relative values are determined from absorption spectra shown in a)-b). The energy difference of $S_1S_0(v_0)$ and $^1[TT]$ with ca. 1000 cm$^{-1}$ is based on previously reported calculations.$^{35}$ d) Spectral correlation maps of the TAP dimer obtain via conventional 2DES (without AP). The population times T are given above the maps. e) Transient 2DES correlation maps using AP and 2DES sequence for a constant population time (T = 150 fs). The individual actinic pump delays $T_{AP}$ are given above the maps. Contours (and lines) are given at a 5% interval.*

It appears to be slightly accelerated compared to excitation of the absorption maximum of the particular TAP dimer ($\lambda_{exc}$ = 688 nm, with $\tau_{iSF}$ = 92 ps),[35] displaying minor excitation-energy dependence. Within the experimentally accessible population-time window (T ≈ 400 ps), no further spectral evolution of the $^1$[TT]-ESA can be discerned from the 2DES maps, which is expected considering the nanosecond-scale lifetime of $^1$[TT] in these dimers.[35]

The evolution of the electronic excited states can be closely followed after the excitation with the AP at the $\lambda_{S0(1)}$-band by keeping a constant population time (T = 150 fs) and varying the actinic pump delay $T_{AP}$ (Figure 2e). Due to the higher-order nonlinearity probed by transient 2DES, a larger prominence of peaks compared to the conventional 2DES experiments (Figure 2d) can be observed. As expected, the main response of the $S_1S_0$ is now located along the diagonal around {590|590 nm} at small $T_{AP}$, matching model descriptions in Figure 1. A clear cross-peak located around {555|590 nm} is also visible, which is related to the excitation of higher vibrational levels of the electronic transition $S_1S_0 \rightarrow S_nS_0(v_x, x>0)$ (see Supplementary material for more details). They overlap with cross-correlations to triplet-related states, the latter of which can also be attributed to the elevated amplitudes around {590|550 nm}. The singlet features decay with the time constant of 80±2 ps, concomitant with the rise of diagonal $^1$[TT]-features at {550|550 nm}, in line with the results from TA after excitation of the $\lambda_{S0(1)}$-band.[35] Lasting cross-peak features can be observed during the evolution from the singlet to the correlated triplet pair, even after the $S_1S_0$ is fully converted (e.g., at $T_{AP}$ = 300 ps, see Figure 2e), indicative of remaining interaction between the adiabatic states. We note that no other major features can be discerned, which shall be elaborated in more detail later.

## High-order Kinetics: Restarting Singlet Fission

The transient 2DES results shown in Figure 3 focus on the evolution of excited states after the second excitation (with the first two pulses $E_1$ and $E_2$) of the 2DES sequence. In turn, we can select the initial state of the system with a constant actinic-pump delay $T_{AP}$ and varying population time T. The majority of the excited-state population occupies the $S_1S_0$ for short $T_{AP}$ of 5 ps, which is fully converted to $^1[TT]$ at $T_{AP}$ = 400 ps. The same $S_1S_0$ features are found in Figure 3a at $T_{AP}$ = 5 ps and T = 0.1 ps, as is the case in Figure 2e where a similar combination of delays is employed. The $S_1S_0$ features decay, now giving rise to the $^1[TT]$ response located in the cross-peak region around {590|550 nm}. The correlated evolution of both states, indicative of iSF, is shown in Figure 3b. An ultrafast evolution (82±5 fs) is also observed in the kinetic trace of the $S_1S_0$ at {590|590 nm}, which is not observed in other regions of the 2D maps. It can be rationalized by the excitation of the 2DES pulse sequence that drives a resonant transition of $S_1S_0$ to an energetically higher electronic state $S_nS_0$, which in turn decays back to the $S_1S_0$ via internal conversion (IC) within that ultrafast time scale. Once $S_1S_0$ is repopulated, singlet fission proceeds, leading to the conversion of the excited singlet state into the multiexcitonic triplet state within the time scale of 78±5 ps (Figure 3c), with $^1[TT]$ state being now correlated to the excitation of $S_1S_0$.

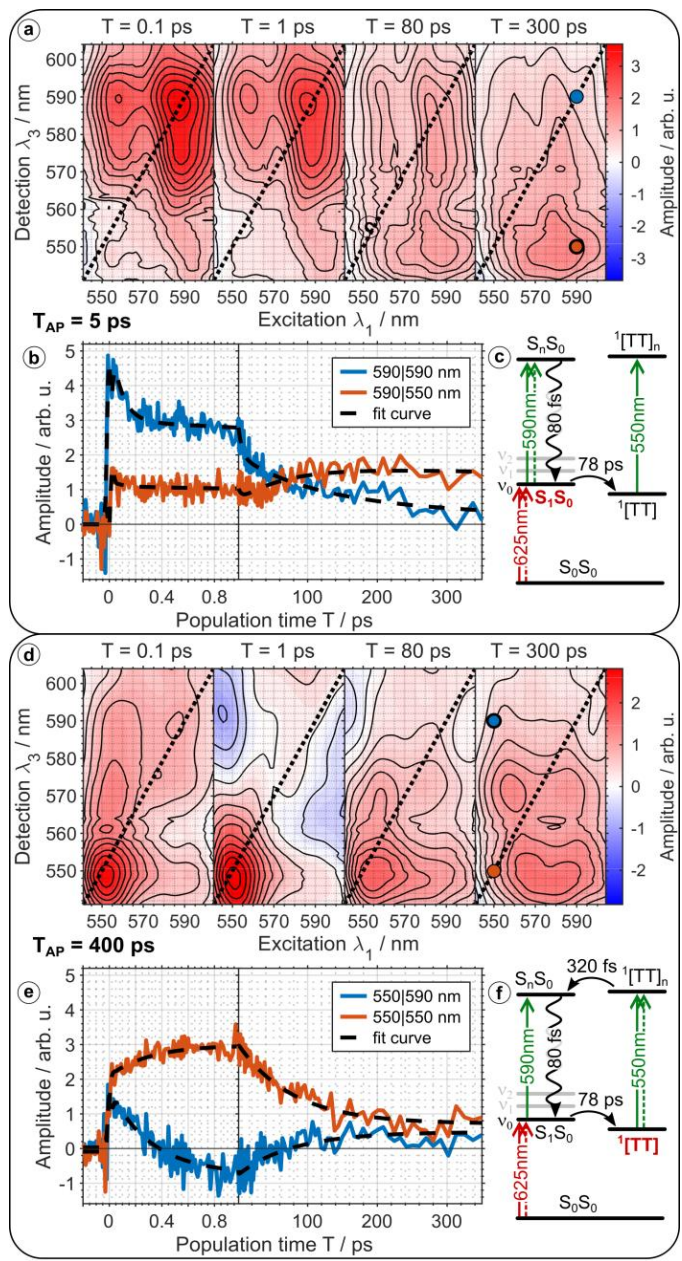

*Figure 3*: Higher-order kinetics for actinic pump delays of a)-c) $T_{AP}$ = 5 ps and d)-f) $T_{AP}$ = 400 ps. a) and d) show transient 2D-correlation maps for four selected population times T (indicated above the maps). Black contour lines are drawn at a 10% interval. b) and e) show selected kinetic traces along T for points indicated in the legend. The selected points are also indicated in the transient 2D correlation maps at T = 300 ps. c) Example diagram for the evolution of the cross-peak at 590|550 nm. f) Example diagram for the evolution of the cross-peak at 550|590 nm. In both cases, c) and f), the energy state populated by the actinic pump at 625 nm is highlighted in a red color.

The excited-state evolution starting from $^1$[TT] shown in Figure 3d ($T_{AP}$ = 400 ps) is evident by the distinct diagonal $^1$[TT] feature at {550|550 nm} and almost non-existing $S_1S_0$ features at {590|590 nm} at early T's (= 0.1 ps). Repopulation of the $S_1S_0$ from excitation of the $^1$[TT] can be

observed by the emergence of a negative cross-peak feature at {550|590 nm} within a timescale of 320±10 fs, as found in the transient 2DES maps (T = 1 ps) and corresponding kinetic traces (Figure 3d,e). The emergence of $S_1S_0$ ESA, now correlated to the excitation from $^1$[TT], can be described in terms of the model in Figure 3f. Initially, the 2DES excitation leads to the population of energetically higher $^1$[TT]$_n$ states, leading to a bleach of $^1$[TT] at {550|550 nm}, and converts to $S_1S_0$ via $S_nS_0$. A definitive involvement of $S_nS_0$ is not evident from the spectra alone, as its spectral features are not visible within the spectral window of the 2DES probe pulse. Nevertheless, based on the quantitative conversion of $^1$[TT]$_n$ to $S_1S_0$, evident by the non-decaying, diagonal $^1$[TT] bleach signal for T < 1 ps (Figure 3e), a direct IC to $S_1S_0$ can be excluded. If it was the case, IC to the initial $^1$[TT] would be expected. Once the $S_1S_0$ is repopulated, the whole iSF process in the dimer restarts, leading to the recovery of $^1$[TT] bleach-features ({550|550 nm}) and dissipation of negative $S_1S_0$ ESA({550|590 nm}) within the characteristic timescale of iSF in this dimer (*vide supra*). Again, lasting correlations remain, even after $^1$[TT] is fully restored, as seen in Figure 3d (T = 300 ps), which shares similarities to the correlation maps shown in Figure 3a ($T_{AP}$ = 5 ps, T = 300 ps) and other actinic pump delays ($T_{AP}$ = 100 ps, see Supplementary material).

## 4. Discussion

The dominant contributions of the intramolecular singlet fission process have been detailed in light of repopulation of $S_1S_0$ from high-energy states that are accessed by the 2DES pulse sequence. Such a circular nature of dynamics after excitation of either $S_1S_0$ or $^1[TT]$ has not yet been reported in literature. In contrast, an earlier work investigating high-order nonlinear responses of bulk, monomeric TAP reported a different outcome when exciting $T_1$ to high-energy triplet states ($T_{2-3}$).[99] No repopulation of singlet states was conceived, but rather, ultrafast IC (<100 fs) back to $T_1$ was observed. This discrepancy in the observed dynamics arises from the differences (in energy and multiplicity) of the triplet-related states $^1[TT]$ and $T_1$. First, considering the multiexcitonic nature of $^1[TT]$, it possesses approximately twice the energy of $T_1$.[4] In turn, the high-energy state ($^1[TT]_n$) populated by resonant excitation of $^1[TT]$ is higher in energy compared to $S_1S_0$, facilitating downhill relaxation, which is not necessarily the case for $T_{2-3}$. Second, the ultrafast nature of the repopulation process points towards spin-conserving IC (instead of intersystem crossing) as underlying mechanism. As a result, we have experimental evidence of the correlated net-zero spin of $^1[TT]/^1[TT]_n$ at longer population times (T > 400 ps) using an all-optical method, corroborating our conclusions from earlier studies,[35-36] and experimental evidence of the different spin-correlated states $^x[TT]$ (x = 1,3,5) using time-resolved electron paramagnetic resonance.[8, 43, 100-101]

The population kinetics of the singlet- and triplet-related states (Figure 2,3) overlap with the cross-peaks related to the interaction between states, obscuring their extraction to quantify relevant coupling parameters. To overcome this, we simulated the high-order responses of the spiro-conjugated TAP dimers for all time-delay combinations (T and $T_{AP}$) of the experiment, shown in Figure 4. A simple model that accounts only for the population kinetics (Figure 3f) properly describes the population-related peaks (see section 3), but fails in explaining the cross-peaks related to the interaction of $S_1S_0$ and $^1[TT]$ (Figure 4 a-b, see also Supplementary material).

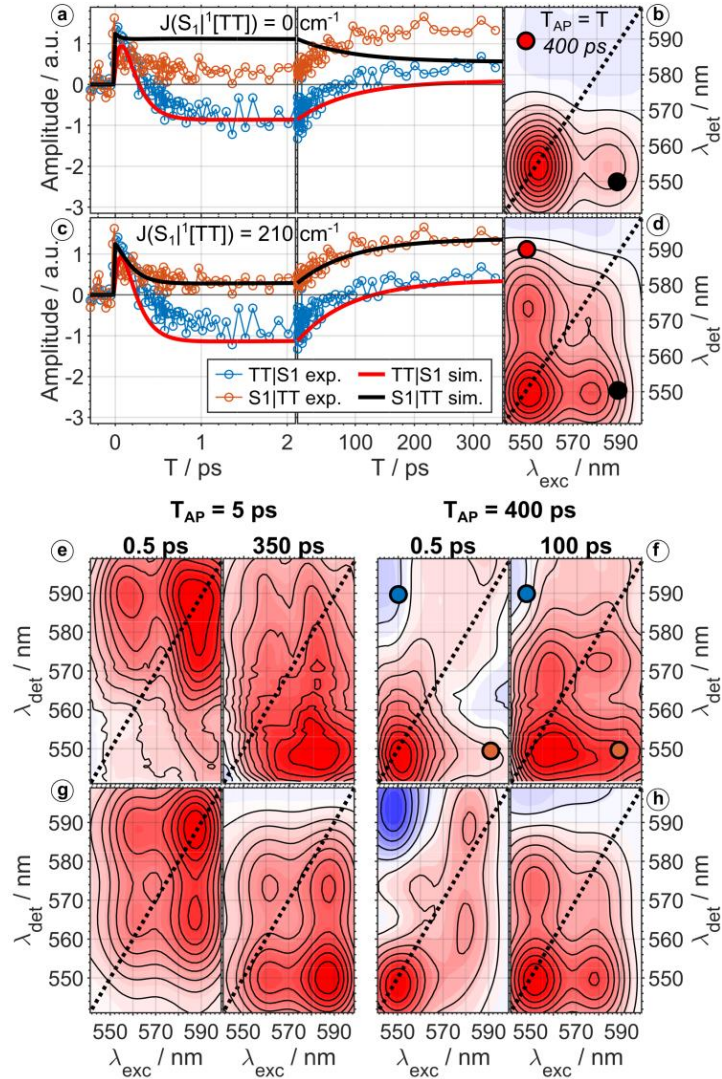

*Figure 4*: Comparison of simulated and experimental transient 2DES correlation maps, a) and c), show kinetic traces of points indicated in the maps of f) and b),d) for T = 0.5 ps, respectively for comparison of experimental data with simulations. The data in b) was simulated for zero off-diagonal coupling element, with a representative map at $T_{AP} = T = 400$ ps shown in b) (see Figure 3d for experimental data at these delays). Accordingly, data in c) was simulated for $J(S_1S_0/^1[TT]) = 210$ cm$^{-1}$ with a representative map at the same delays as in d). Experimental correlation maps for an actinic pump delay $T_{AP}$ of 5 and 400 ps are given in e) and f), respectively, with simulated maps at the same $T_{AP}$'s given below in g) and h), respectively. The corresponding population times T are given above the experimental transient 2DES maps. All maps are normalized to 1 with contour lines at 10% intervals.

An accurate resemblance of simulation with experimental data (Figure 4 c-h) is achieved by inclusion of direct exchange couplings $J(S_1S_0|^1[TT])$ and $J(S_nS_0/^1[TT]_n)$. The addition of super-exchange coupling via CT states did not yield accurate representations of our experimental data. In turn, we can completely rule out super-exchange coupling via CT states mediating the iSF of the spiro-conjugated dimers, as predicted in our earlier work.[35] We can explain this phenomenon based on the comparatively high energy of the CT states in the TAP spiro-dimers, which decreases its relevance to the iSF process. Higher contributions of CT might be the reason for the enhanced iSF kinetics after excitation of higher vibrational eigenstates of $S_1S_0$ ($v_3$), as observed in the conventional 2DES experiment (Figure 2). This is no longer the case for transient 2DES, for which vibronic transitions with lower energies are driven by the AP. For this case, direct SF is the main pathway. The good agreement between simulations and experimental results (Figure 4 and Supplementary material) lets us quantify the value of the direct exchange coupling of $J(S_1S_0|^1[TT]) = 210$ cm$^{-1}$ mediating iSF in the dimers. The magnitude of this coupling term is rather small, which is consistent with earlier theoretical results.[35] The small coupling is compounded by a comparatively large energy separation of $S_1S_0$ and $^1[TT]$ with ($\Delta E \approx 1000$ cm$^{-1}$) resulting in the rather slow conversion time of $S_1S_0$ to $^1[TT]$, considering Fermi's golden rule.[4] Our simulations also suggest a non-zero coupling element of high-energy states ($J(S_nS_0|^1[TT]_n = 110$ cm$^{-1}$). However, in this case, a smaller energy separation (see Figure 2c) leads to a faster transition (*vide supra*). More importantly, the result of this mixed state is found in the transient 2DES maps at late T-values (for any $T_{AP}$): The apparent lasting signatures found at {550-590|575 nm} are the result of the mixed electronic states, where excitation of the singlet- or triplet-related states can cohere with the mixed electronic states.

## 5. Conclusion

In conclusion, singlet fission in the spiro-conjugated homodimers presented here proceeds via a direct conversion mechanism, as determined experimentally by transient 2DES. The direct SF is mediated by weak coupling (210 cm$^{-1}$) of singlet and correlated triplet pair leading to the observed 80 ps iSF dynamics. The remarkable selectivity of transient 2DES to signals stemming from excited states also pointed towards mixing of high-energy states ($S_nS_0$ and $^1[TT]_n$) that are made accessible by the additional excitation pulses. This leads to a cascade of ultrafast IC (<350 fs) back to the $S_1S_0$, restarting SF even at later stages of the process, which is the dominant contribution to the high-order stimulated signal in the investigated dimers. By means of semi-quantum mechanical simulations and the remarkable agreement with experimental data over a vast combination of delays, we were able to extract the weak exchange coupling relevant for SF. This feat of direct observation of the excited-state coupling element has been presented here for the first time and will lead the way to a better understanding of SF once applied to a multitude of SF sensitizers.

**Acknowledgements**

This research was funded by the Deutsche Forschungsgemeinschaft (DFG) via the Sonderforschungsbereich SFB 1249-A01 and SFB 1249-B04.

**Authors disclosure statement**

Authors declare no competing financial interests.